# Multiscale Experimental Evidence of a Transitory State in the HDA-LDA Transition in Alumino-sodo-silicate Glasses

**Antoine Cornet[1][*], Thierry Deschamps[1], Sylvie Le Floch[1], Dominique de Ligny[2] and Christine Martinet[1]**

[1] *Institut Lumière Matière UMR CNRS 5306, Université Claude Bernard Lyon 1, CNRS, F-69622 Villeurbanne, France*

[2] *Institute of Glass and Ceramics, Department of Materials Science and Engineering, Friedrich Alexander University Erlangen Nuremberg, Martensstr. 5, 91058 Erlangen, Germany*

*corresponding authors: antoine.cornet.work@gmail.com

## Abstract

The structural evolution with temperature of pure silica ($SiO_2$), sodium-silicate ($5Na_2O$-$95SiO_2$, $10Na_2O$-$90SiO_2$ and $25Na_2O$-$75SiO_2$) and albite ($15Na_2O$-$15Al_2O_3$-$75SiO_2$) glasses previously densified from hot compression is monitored with a combination of small and wide angle x-ray scattering. Transient scattering maxima in the length scales associated with both techniques indicate the presence of a transitory state, suggesting the classification of the initial and final states as distinct glass states in the context of polyamorphism. In all glasses, a transient intensity peak in the small angle scattering is observed, consistent with the nucleation of a new amorphous state. Based on a comparison with the phenomenology observed in supercooled and glassy water, we propose that this structural evolution provide a strong evidence of the glass polyamorphism as the consequence of the existence of an underlying but thermodynamically defined liquid-liquid transition, independent on the polymerization degree of the system.

## 1 - Introduction

A typical phase diagram includes the gaseous and liquid states in addition to a multitude of crystalline solid phases, where the transitions between these different phases are triggered by an external parameter such as pressure or temperature. Polymorphism, that is the existence of distinct crystalline phases, takes place even in the simplest compounds such as hydrogen [1,2]. In the amorphous liquid state, the possibility of the existence of distinct phases, a concept dubbed as polyamorphism, was initially invoked to explain the presence of a maximum in the melting curve of few chemical compounds [3]. Later on, polyamorphism gained importance as a potential explanation for the unusual physical behaviour of water [4], and its observation in various systems such as metallic [5–7], chalcogenide [8,9], oxides [10,11], molecular [12] and ionic [13] liquids suggested a widespread occurrence in the liquid state. The locus of the liquid-liquid transition (LLT) lines and the associated critical point is often localized deep below the melting temperature in the undercooled regime [14], and polyamorphism can be masked by crystallization, or can take place in the out-of-equilibrium glass state [15–17]. In the latter, the signature of an well-defined first order transition in the liquid state can become kinetically slowed down, or even arrested, making a proper assignment of polyamorphism a difficult task.

In silica glass, which is the archetypical glass and forms the backbone of magmatic and technologically relevant compositions, polyamorphism has been used to rationalize the large density and structural differences obtained after a compression above the elastic limit at 9 GPa at room temperature [18–21]. Largely based on the phenomenology taking place in amorphous water ice, the pristine low density and densified high density states have been called HDA and LDA for High and Low Density Amorphous sates, and it has been suggested that these two states are distinct glassy states corresponding to a case of polyamorphism [22–24]. However, the different structural changes taking place during the compression and leading to the final densification of the glass span over a pressure too wide to assess unambiguously a case of polyamorphism and the nature of the potential LDA→HDA transition [25].

One way to circumvent the problem in the assessment of the LDA→HDA transition is to start with densified samples and induce the relaxation with high temperature at atmospheric pressure, that is the potential HDA→LDA transition. This approach coupled to a structural monitoring shows a recovery of the pristine properties in a short temperature range (or associated with short relaxation times at fixed temperature) for a variety of oxide glasses including silica [26–30]. The same approach with a structural monitoring based on a combination of Raman spectroscopy and x-ray scattering allowed us to demonstrate the existence of a transitory state characterized by a maximum in the density fluctuations at various length scales within the Intermediate Range Order (IRO) of the glass, that we associated with an activated state [31,32]. The presence of this activated state indicates that the high and low density glass states correspond to distinct energy minima of the potential energy landscape (PEL), that is in the potential energy of the glass with respect its structural degrees of freedom [33,34], and provides an argument for the proper definition of HDA and LDA as distinct states and for the occurrence of a pressure induced polyamorphism in silica.

The aim of the present study is to test the extension of this polyamorphism to binary and ternary alumino-sodosilicate compositions to characterize the link between polyamorphism, polymerization and network modifiers. The progressive addition of $Na_2O$ to $SiO_2$ up to 25% $Na_2O$ induces a gradient of polymerization in the corresponding samples. Adding $Al_2O_3$ to reach an equimolar proportion of Na and Al allows in turn to obtain a fully polymerized sample, and therefore to investigate selectively the effect of network modifiers and polymerization on polyamorphism. All samples were densified at or close to their maximum densification ratios from hot compression, and subsequently heated at atmospheric pressure with a structural monitoring performed by simultaneous Small and Wide Angle X-Ray Scattering (SAXS/WAXS). The results show the apparition of a transitory state in the corresponding structural length scales in all samples, and unveil a phenomenology that resembles that of the liquid-liquid (glass/glass) transition in water.

## 2 – Experimental

*2-1 - Densification at high pressure-high temperature*

For a thorough characterization of the polymerization degree on the pressure-driven polyamorphism in silicates glasses, 5 samples with different compositions were synthetized and further densified. The full list of samples/compositions is available in the table I. The starting sample is pure $SiO_2$, which is fully polymerized. From this, the degree of polymerization is gradually decreased by introducing sodium oxide with molar ratios of 5%, 10% and 25%. The obtained glasses are labelled NSX, indicating the $SiO_2$ proportion ($Na_2O$-$XSiO_2$) and leading to NS19, NS9 and NS3. The Sodium atoms act as network modifiers: the relative positive charge bared by $Na^+$ ions is compensated by the breaking of a Si-O-Si bond in its immediate proximity, leading to the formation of non-bridging oxygens $Si-O^-$, hereby decreasing the overall polymerization. Adding Aluminium oxide leads to the ternary $SiO_2$-$Na_2O$-$Al_2O_3$ system in which, as long as $X_{Al/Na+Al}<0.5$, the $Al^{3+}$ ions makes tetracoordinated $AlO_4$ units, the negative charge being compensated by a nearby $Na^+$ ion [35]. The aluminium acts as a network former and the polymerization increases again. The fifth sample of study is therefore glassy albite, which presents an equal amount of Na and Al and is fully polymerized again.

The NS19 and NS9 samples were prepared from commercial Sigma-Aldrich high purity $SiO_2$, $Na_2CO_3$ powder reactants. Before melting, the powders were mixed into Pt crucible and annealed at 300 °C during two hours and subsequently at 800 °C during eight hours in a dedicated furnace for water removal and decarbonation. The crucible was then covered and placed in a high-temperature furnace for the melting. From the phase diagrams [36], the melting temperatures were chosen as 1650 °C (NS19) and 1700 °C (NS9) to ensure that the homogeneous liquid phase is reached. The melting temperatures were then maintained for a sufficient time to ensure bubble free glasses. Finally, the melts were quenched by immersing the bottom of the crucible into water. ICP-OES spectroscopy results show that the final compositions are the same that the nominal compositions within 1% errors.

The NS3 and albite glasses was prepared from $Al_2O_3$, $Na_2CO_3$ and $SiO_2$ powders following the protocol described in ref [37]. The initial powders were dried (1000 °C for $SiO_2$ and $Al_2O_3$, 350 °C for $Na_2CO_3$) and crushed in a mortar in ethanol. Decarbonation was done by heating up to 1500 °C, and subsequently quenched by immerging the bottom of the platinum crucible in water. Repeated heating-quenching-grinding ensured homogeneity, followed by a long annealing to obtain a bubble free glass. The obtained densities match that reported in previous synthesis of binary silicate glasses [38,39].

| Chemical composition | Density (g.cm$^{-3}$) | Label |
|---|---|---|
| $SiO_2$ | 2.2 ± 0.01 | $SiO_2$ |
| $5Na_2O – 95SiO_2$ | 2.24 ± 0.01 | NS19 |
| $10Na_2O – 90SiO_2$ | 2.29 ± 0.01 | NS9 |
| $25Na_2O – 75SiO_2$ | 2.44 ± 0.01 | NS3 |
| $12.5Na_2O – 12.5Al_2O_3 – 75SiO_2$ | 2.35 ± 0.01 | Albite |

*Table 1 - Chemical compositions, densities and denominations of the different glasses. The densities were estimated via buoyancy measurements in toluene at room temperature.*

Binary $SiO_2$-$Na_2O$ glasses with 20% to 25% (mol) of $Na_2O$ are known to undergoe phase separation when heated, or when the quenching is not fast enough[40–43]. However, it will be shown that the transient state previously observed in pure silica can be unambiguously distinguished from the chemical ordering due to phase separation in the NS19 and NS9 glasses.

Cylinders (height: 4 mm; length: 6 mm) were cut from the raw pellets and densified in a Belt type press apparatus. The same procedure used for the compression of the previous silica samples was reiterated [44]. Samples were first compressed up to the nominal pressure at 10 bar/minute, using pyrophyllite as pressure transmitting medium. Temperature was then set to the desired value with a soft heating ramp (≈1 °C/s), and maintained during ten minutes (twenty for the albite glass). Finally, the temperature is shut down and the pressure decreased after this quenching. The maximum pressure and temperature reached during the compression process are summarized in table 2 for each composition. To allow for comparison, the maximum pressure reached for all the samples is the same than for the already studied silica samples: 5 GPa. At atmospheric pressure, the phase separation within the NS19 and NS9 glasses starts from 500 °C [40,41]. Although this temperature might change with pressure, we used it as reference, and performed our compression at 300 °C.

The densities of the recovered samples, estimated via the buoyancy method in toluene, are summarized in table II along with the corresponding densification ratios. The densification ratios obtained after compression are around 15% for all glasses, except for the NS3 sample, where it reaches 8.2%. These values can be compared to the empirical maximum values to characterize the efficiency of the densification process. Using the known Poisson ratios for silica [18], albite [45] and $Na_2O$-$SiO_2$ [42] glasses, the maximum densification ratio can be estimated from the calibration $(\Delta\rho/\rho)_{MAX} = 150 \times exp(-13 \times \nu)$ established by Rouxel et al. [18]. This gives maximum densification ratios of 21%, 16.7%, 14.5%, 7.1%, 13.7% for the SiO$_2$, NS19, NS9, NS3 and albite glasses respectively. For each recovered sample except silica, the densification ratio obtained are, within the measurement's uncertainty, equals to the maximum value given from the empirical calibration. Thus, the parameters chosen for the compressions allow complete densification of the samples. In the case of silica, the densification curves show similar trend between 900 °C and 1100 °C, suggesting that higher pressures are needed to achieve the maximum densification ratio [21], which was not achievable in the press used in this study.

| Label | Pressure (GPa) | Temperature (°C) | Density (g.cm$^{-3}$) | Densification ratio $\frac{\rho-\rho_0}{\rho_0}$ (%) |
|---|---|---|---|---|
| SiO$_2$ | 5 | 1020 | 2.56 ± 0.02 | 16.5 ± 1.4 |
| NS19 | 5 | 300 | 2.58 ± 0.03 | 15.2 ± 1.9 |
| NS9 | 5 | 300 | 2.65 ± 0.03 | 15.6 ± 1.8 |
| NS3 | 5 | 500 | 2.64 ± 0.02 | 8.2 ± 1.3 |
| albite | 5 | 600 | 2.71 ± 0.01 | 14.3 ± 0.9 |

*Table 2 - compression parameters, recovered densities and denomination of the different samples. The densities were estimated via buoyancy measurements in toluene at room temperature.*

*2-2 – X-ray scattering at high temperature*

Combined in-situ SAXS and WAXS experiments were performed at the BM26 beamline of the ESRF (Grenoble, France), using a monochromatic beam operating at 12 KeV (1.033 Å), using the same setup described previously [44] for the pure silica glass experiments. The small-angle diffusion intensity is recorded with a Pilatus 1M detector placed 1.4 meter away from the sample, with the transmitted x-rays traveling through a primary vacuum, for a spectral detection window ranging from 0.12 nm$^{-1}$ to 7 nm$^{-1}$. Wide angle diffraction signal was recorded with a Pilatus 300K detector placed at approximately 15 cm to the sample, only allowing limited monitoring of the structure factor over 6 or 8 to 50 nm$^{-1}$, depending on the experimental session. Closer sample-WAXS detector arrangement was not possible because of the SAXS vacuum pipes. Temperature control was performed by an iteration of the ESRF custom-made microtomography furnace [46] equipped with Kapton windows, using a 10 °C/min ramp rate. All data correspond to 60s scans, showing a good compromise between time resolution and signal/noise ratio. The empty furnace signal was used as a single background for all scans, and each scan was corrected for the incoming flux, measured just upstream from the sample position with an ionization chamber.

Geometrical calibration of the SAXS and WAXS detectors was done from the scattering of silver behenate and α-Al$_2$O$_3$ respectively. Azimuthal integration was performed using the Pylatus software originally developed at the BM01 beamline of the ESRF [47,48].

## 3 – Results

The evolution of the small-angle scattering from the densified SiO$_2$ sample upon heating is visible on the left part of the figure 1. Specifically, the left panel contains all curves obtained during the heating of the densified sample and the subsequent annealing at constant temperature, with the first scattering curve at the lowest temperature in dark blue, and the last acquired one in dark red. It is clear that the evolution of the scattered intensity is complex, and depends on both the considered scattering vector/length scale and the temperature/time. For this reason, we chose to represent the evolution of the relative intensity with respect to the first, reference curve, as a function of time. This is done in the right panel of the figure 1, along with the temperature evolution plotted on the same temporal axis.

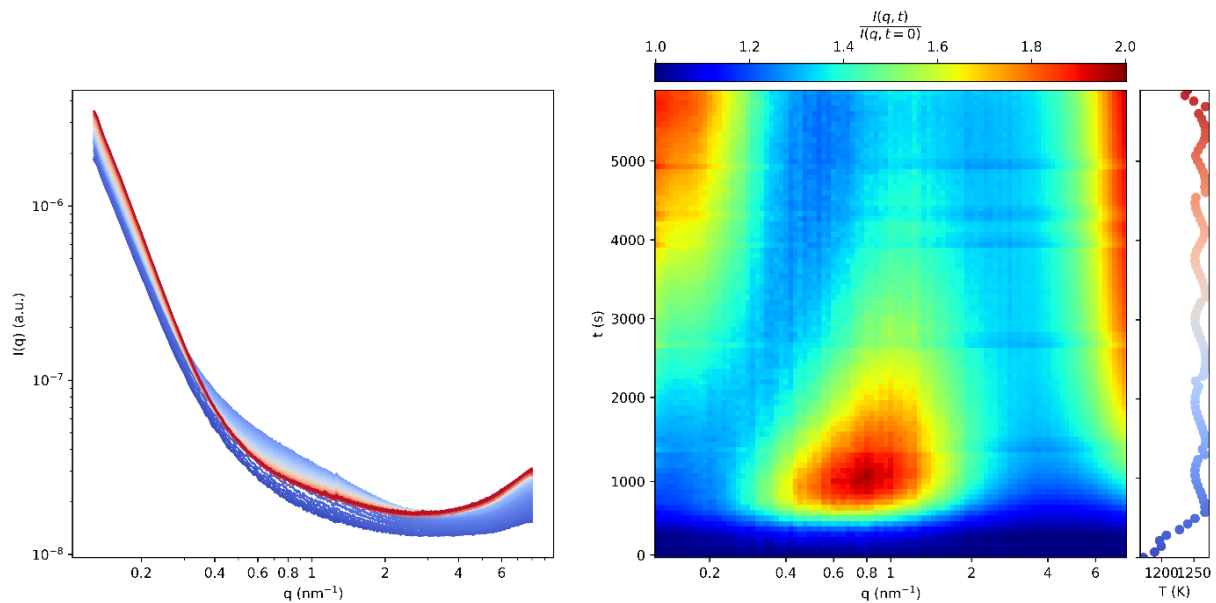

*Figure 1 - Left : SAXS scattering curves of a silica sample previously densified at 5 GPa and 1020°C. Right: Relative intensity map $I(q,t)/I(q,t=0)$, together with the tempearture evolution. The color coding of the scattering curves corresponds to that of the temperature evolution.*

Despite the relatively high starting temperature for this sample due to a beam dump on heating, this representation highlights clearly the presence of a transitory state in the annealing of silica densified from hot compression. This transitory state is observed at scattering vectors ranging from 0.3 to 3 nm$^{-1}$ with a maximum at 0.8 nm$^{-1}$, and expands in time over 4000 s at the annealing temperature of 1260±8K. In addition to this transitory state, we observe an increase of the scattering intensity at the largest and smallest scattering vectors, which can be attributed to a shift or an increase of the FSDP and an increase in the liquid-like density fluctuations. The latter increase of the density fluctuations is not the common trend in pristine SiO$_2$, where a decrease of the scattered intensity is visible upon relaxation [49]. This highlights the different structures between the reference and hot compressed silica, as it has been shown already at smaller length scales [50].

The small angle scattering relative intensity maps of NS19, NS9, NS3 and albite hot compressed glasses upon heating and annealing are displayed on the figure 2. All glasses show a transient state in the low-$q$ scattering during the heating from room temperature, with two qualitatively different trends for the low (NS9 and NS19) and high (NS3 and albite) Na content. The NS19 and NS9 glasses present a concomitant reduced and increased scattering intensity at lower and larger q-values respectively, with an increase that can extend to the beginning of the wide-angle regime, up to 5 nm$^{-1}$ for the NS19 glass. The region of increased scattering peaks over a relatively narrow range in time (1360-2200s for NS19, 1025-1850s 1024-1863 for NS9), and consequently in temperature (526-666K for NS19, 473-612K for NS9), while the reduced scattering at lower $q$ starts at similar times/temperatures but expands to higher temperature, even to the annealing temperature in the case of NS19. While the intensity rise is expected given the similar trend in pure silica, the decrease in scattering at lower $q$ can be attributed to a continuous structural evolution: this disappearance is not transitory, as it persists well after the vanishing of the local intensity rise. When interpreting the intensity signature of the transitory state alone, the apparition of a single peak can be interpreted in the Ornstein-Zernicke theory as the structural signature of an arrangement of scatterer, whose shape is ideally Lorentzian [51,52]. It reveals the presence of correlations in the electronic density associated with a length scale of $d=2\pi/q\approx3$nm for the NS19 and NS9 glasses, and 7.5 nm in the case of silica, yielding a length scale particularly large for fully polymerized oxide glasses. At the $q$ value of where the transient state is the strongest, the local intensity rise reaches 0.66±0.02, 0.06±0.02 and 0.25±0.02 in pure silica and in the NS19 and NS9 samples respectively (in percentage of the reference intensity, figure S5), showing that there is no monotonous trend for the strength of the density fluctuations during the relaxation mechanism with the sodium content at this length scale.

The top left high intensity feature on the NS9 scattering map can be attributed to the inevitable phase separation of binary sodo-silicate glasses containing < 20% mol of Na [53]. Indeed, the temperature profile of the intensity rise coincides with the rise of the heterogeneities monitored through the Landau-Placzek ratio in a closely related NS7.3 (non-densified) glass [43] (figure S6). To the best of our knowledge, there is no data available regarding the monitoring of the chemical ordering during the phase separation in the NS19 glass, or in a close composition. Therefore, by similarity with the phenomenology observed in the NS9 glass, we attribute the increase of the scattering intensity from 500s onward and at $q<0.4$ nm$^{-1}$ to the phase separation mechanism.

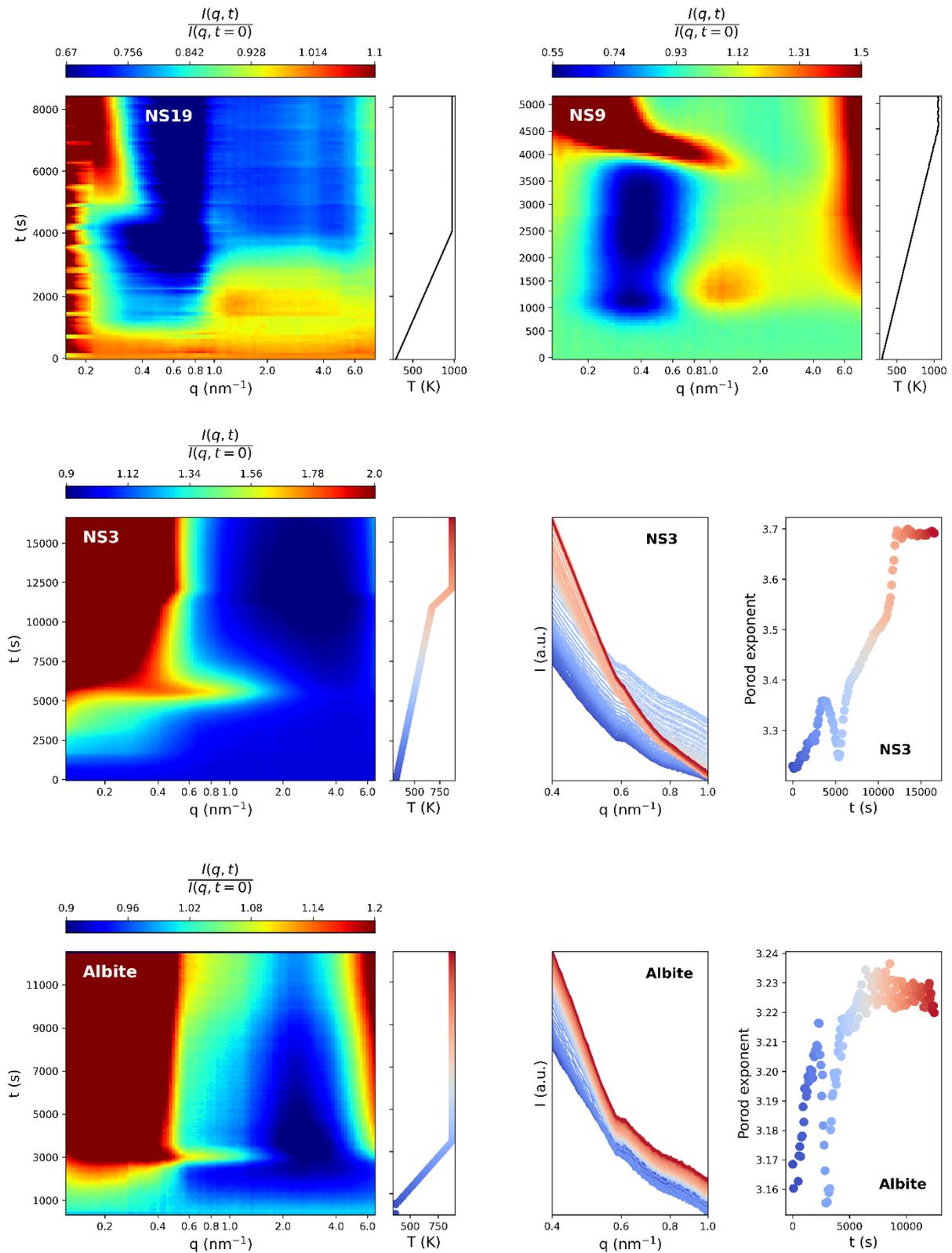

*Figure 2 - SAXS relative intensity maps for the densified NS19 (top left), NS9 (top right), NS3 (center) and albite (bottom) glasses, along with the temperature evolution. Scattering curves and Porod exponent, obtained from a power law fitting between 0.17 and 0.28 nm$^{-1}$ are also plotted for NS3 and albite glasses. For these sampels, the color coding of the scattering curves and Porod exponent corresponds to that of the temperature evolution.*

Here, we must point that it is difficult to obtain a unified, across the scales description of the phase separation from a structural point of view. The different techniques used to probe this phase separation typically probe different length scales: the SAXS intensity provides monitoring over few nm, the Landau-Placzek ratio relies on the Rayleigh scattering of visible light, hence covering the same nm length scale, while imaging typically shows domains on the micron/submicron scale [42,43]. Thus, inconsistencies can appear in the temperature ranges reported for the phase separation across the range compositions and techniques considered. Nevertheless, our results demonstrate that the spinodal decomposition is kinetically arrested at temperatures up to 800K, in line with previous SAXS measurements in a NS6.4 glass (13.5Na$_2$O-86.5SiO$_2$) [54].

A transient state also appears in the intensity evolution of the densified NS3 and albite, although its appearance is qualitatively different from the previous NS19 and NS9 glasses. First, the most prominent feature for both NS3 and albite glasses is a strong and monotonous increase of the intensity at low-$q$, showing increasing liquid-like fluctuations in the electronic densities. While the colour scale is adapted to highlight the transient state, it is worth to note that this increase reaches 70% and 600% of the room temperature intensity for the densified albite and NS3 glasses respectively. However, similar trends, albeit with lower magnitude (20% and 200%) in the non-densified reference glasses (figures S3 and S4) suggest that the effect is at least partly due to a relaxation of the glass upon annealing, and is enhanced with densification process. Overall, it reveals that the hot compression promotes the homogeneity of the glassy network since the density fluctuations giving rise to the scattering increase during the relaxation, as already demonstrated in pure silica glass [50]. A power law modelling of the scattering curves over the restricted $q$-range of 0.17 to 0.28 nm$^{-1}$ yields a Porod exponent ranging from 3.16 to 3.24 for the albite glass, and 3.2 to 3.7 in the NS3 glass. Such values between 3 and 4 demonstrates the lack of well-defined boundaries in the electron density contrast, consistent with the percolating network already identified in molecular dynamic simulations, where high density Na pockets shows smooth boundaries with the probed Na local density [38]. The larger evolution of the Porod exponent in the NS3 glass against that in the albite glass (0.08 vs 0.5 respectively) shows that the sub-structure of the Na atoms is largely stabilized by the charge compensation induced by the nearby Al atoms. To the best of our knowledge, there is no atomistic picture (either by molecular dynamics or from microscopic studies) to identify the homogenous or heterogenous spatial distribution of the Na$^+$ network modifiers when these network modifiers are strictly compensated for by nearby Al$^{3+}$ ions. As such, it is unclear for us whether the pockets and channels of high density modifiers first identified by Greaves [55] are still present at Al/Na=1. Therefore, we can only build on the phenomenological similarity between the relaxation of hot-compressed NS3 and albite glasses at the length scale associated with SAXS to suggest the existence of heterogeneously distributed Na ions even when Al/Na=1. In both cases, the Porod exponent shows a well-defined local maximum in the otherwise continuous increase. This indicates first that the global trend is that the boundary of the Na rich pockets/channels becomes more defined during the relaxation, and that a transient state exists where the opposite process takes place. The corresponding elapsed times for this transient state are 5400s and 3100s for NS3 and albite glasses respectively.

The transient state also appears as a spike from this global low-q intensity rise that expends to larger scattering vectors up to 2 nm$^{-1}$, and seems to peak around 0.4-0.8 nm$^{-1}$. In both glasses, the transient state appears during the heating stage, allowing to define a temperature window for the phenomenon at the rate of $4.1 \times 10^{-2}$ K/s : 4500-7400s (464-557K) for NS3 ; 2300-3400s (669-864K) for albite. For each sample, this corresponds to the range of elapsed times identified from the Porod exponent. This transient state corresponds to a $q$-range where oscillations of a form factor is visible, as shown in figure 2. However, similar oscillations are also visible in the NS19 and NS9 glasses, mostly in the pristine glasses (figures S1 and S2), and the discrepancy in the scattering vectors $q$ between these oscillations and the transient state in these low Na content glasses dissociate the intensity rise to the visible form factor. Based on this $q$ discrepancy, we suggest that the transient state also appears as localized intensity rise in the NS3 and albite glasses, which happen to take place at the same $q$-range as the form factor.

Next, we discuss the structure at the level of the atomic scale, obtained by x-ray diffraction. In this study, we focus on the so-called First Sharp Diffraction Peak (FSDP) as a marker of the Medium Range Order (MRO) of the different glasses, because the Short Range Order (SRO) remains untouched at atmospheric pressure whatever the densification process, as seen from the O-Si-O angle distribution in silica [56].

The region of the FSDP is characterized by multiple components to the main peak, from a pre-peak around q = 9 nm$^{-1}$ to the high-q side shoulder to the main peak at q = 24 nm$^{-1}$. Nevertheless, for all compositions but the NS3 glass, a well-defined asymmetric peak is visible. Here we modelled the top 50% of this peak through the fitting function:

$$I(q) = y_0 + A \frac{1}{1 + e^{-\frac{q - q_c + \omega_1/2}{\omega_2}}} \times \left[ 1 - \frac{1}{1 + e^{-\frac{q - q_c - \omega_1/2}{\omega_3}}} \right]$$

This function does not hold any physical meaning but allows for an analytic description of the scattered intensity with a good description of the asymmetry, visible through the low residuals plotted figure 3. This decision was motivated by the lack of consensus on the structural features responsible for the scattering of the FSDP, which prevents the definition of a physically based model. The analytical description allows for the extraction of the peak width (at half maximum) $\Gamma$ and maximum position $q_1$ with "infinite" q-resolution. In pure silica, the transient state appears as a local maximum in $\Gamma$, while the evolution of $q_1$ is monotonous, so we concentrate in the main text on the monitoring of $\Gamma$. Specifically, we report the evolution of the width $\Gamma_x$ at different percentages of the maximum intensity, from 50% ($\Gamma_{50}$) to 90% ($\Gamma_{90}$), normalized by its value at the beginning of the relaxation and as a function of the elapsed time for the SiO$_2$, NS19, NS9 and albite samples. The corresponding temperature evolution is also reported.

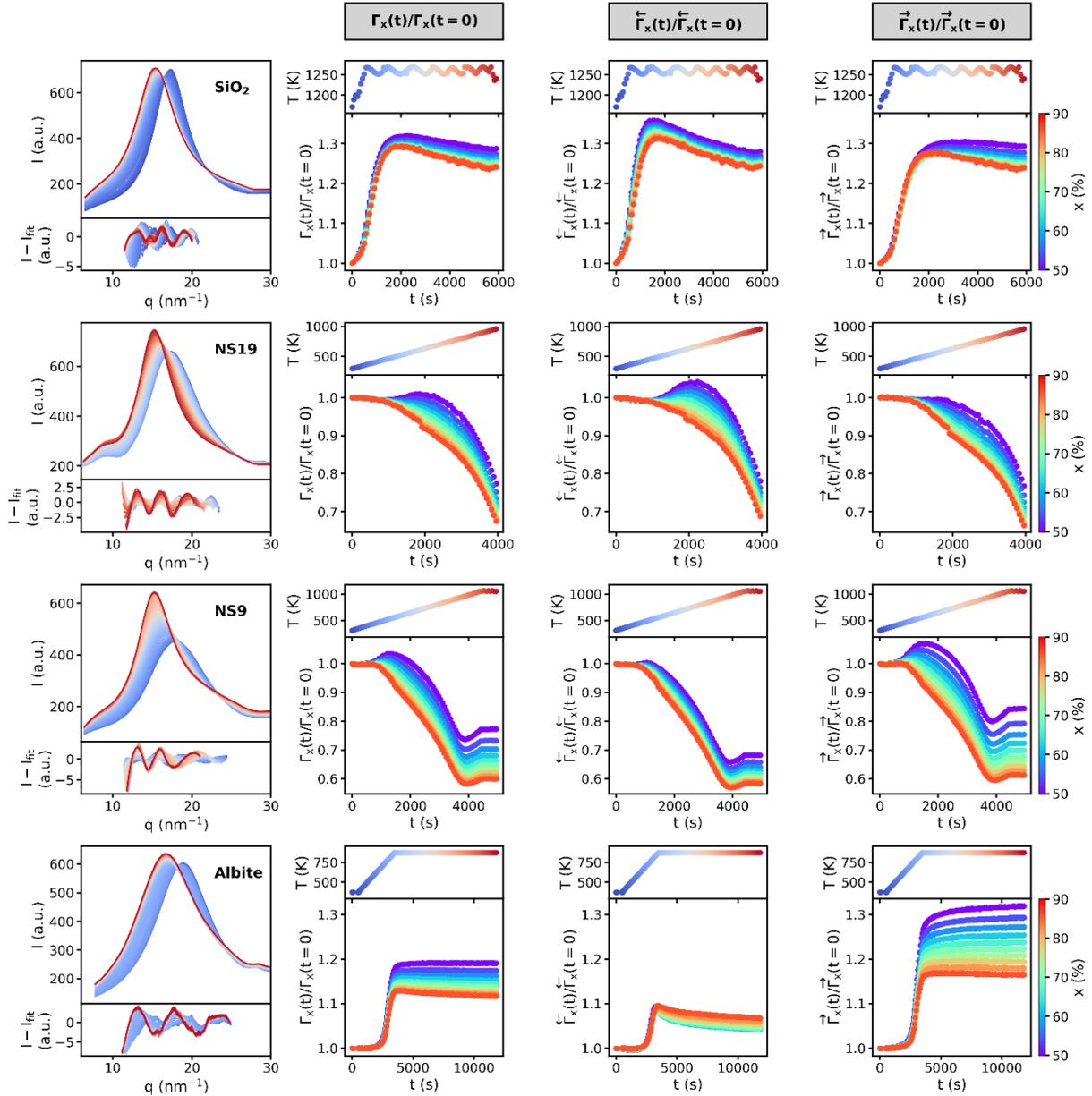

*Figure 3 - WAXS scattering curves (first column) for the SiO$_2$, NS19, NS9 and albite glasses, and residual from the modelling of the top 50% maximum intensity of each curve with the function described in the main text. Corresponding evolution of the relative FSDP full width $\Gamma_x(t)/\Gamma_x(t=0)$ (second column), left $\overleftarrow{\Gamma}_x(t)/\overleftarrow{\Gamma}_x(t=0)$ (third column) and right $\overrightarrow{\Gamma}_x(t)/\overrightarrow{\Gamma}_x(t=0)$ (fourth column) half width at x % percent of the maximum intensity as a function of time. The time axis of the NS19 and NS9 glasses has been cut to avoid the phase separation region. The color coding of the scattering curves and residuals corresponds to that of the temperature evolution.*

For all samples apart from pure silica, we note that the presence of a local maximum in the width $\Gamma_x$ depends on the percentage x of the peak considered, which indicates a complex atomistic mechanism which differs at a function of the length scale. As the width evolution is not homogenous through the peak height, the evolution $\Gamma_x$ implies a local intensity rise or decrease at a certain range of $q$ within the whole peak: the farther x is from 50, the farther the transient intensity change is from the peak maximum. This appears clearly when considering the half width of the FSDP at its low and high $q$ sides, denoted $\overleftarrow{\Gamma_x}$ and $\overrightarrow{\Gamma_x}$ respectively, and plotted in the third and forth columns of the figure 3. The SiO$_2$ goes through a transient state mostly on the low $q$ side of the peak with a local maximum from $\overrightarrow{\Gamma_{50}}$ to $\overrightarrow{\Gamma_{80}}$, while it is strictly contained to $\overleftarrow{\Gamma_x}$ in the NS19 ($\overleftarrow{\Gamma_{50}} \to \overleftarrow{\Gamma_{70}}$) and albite ($\overleftarrow{\Gamma_{50}} \to \overleftarrow{\Gamma_{90}}$) glasses. The NS9 sample contrasts with these results in that the transient width effect is observed in the high $q$ side of the FSDP, from $\overrightarrow{\Gamma_{65}}$ to $\overrightarrow{\Gamma_{50}}$. This indicates that the structural mechanisms at play depend on the degree of polymerization and not on the Na content, with similar observations on the highly polymerized SiO$_2$, NS19 and albite samples, as opposed to the less polymerized NS9 glass. A possible reason for the local intensity rise/decline lies in the minor peaks at the edges of the FSDP, visible at q = 9 nm$^{-1}$ and q = 24 nm$^{-1}$. However, our attempts at a deconvolution did not provide consistent results through the full relaxation, mostly due to the complex shape of the FSDP. We note that the transient width effect in pure SiO$_2$ extends to the high $q$ side of the FSDP, 10 nm$^{-1}$ away from the low $q$ peak. This same low $q$ peak is more decoupled from the FSDP in the NS19 glass, allowing for the extraction of its width (at half maximum): 0.7 nm$^{-1}$. If the width of the peak in NS19 is transposable to silica, then it cannot be responsible for the rise of the transient width effect in SiO$_2$ and, by extension, to the other compositions. The evolution of the position $q_1$ of the peak maximum is available in the figure S7. It's derivative with the elapsed time indicates a maximum rate for the structural change that precedes the local maximum seen in the width.

The FSDP of the NS3 glass shows two strong contributions, visible in figure 4, consistent with previous x-ray scattering observations [57]. The apparition of a second contribution to the FSDP in binary (borate) glasses was suggested to arise from an arrangement of voids centred around the network modifiers [58], although the lack of a consensual interpretation of the FSDP precludes a definitive assignment. Nevertheless, the general evolution of the structure factor of xSiO$_2$-(1-x)Na$_2$O glasses shows that the low $q$ component is silica like, while the high $q$ contribution is representative of the structure of binary glasses with high Na content [59]. The presence of both peaks is consistent with the partition of the overall structure in distinct Si rich and Na rich regions, each associated with a FSDP component. It is important to note that this is not equivalent to the binodal or spinodal phase separation in binary melts and glasses mentioned above, as compositions with more than 25% (molar) Na$_2$O remain monophasic [60], but to the well-known Na-rich pockets and/or channels developing in the SiO$_2$ rich network [55,61]. As such, the high $q$ contribution of the FSDP might could be considered as a proxy for the structural evolution of the densified binary glasses with high Na content.

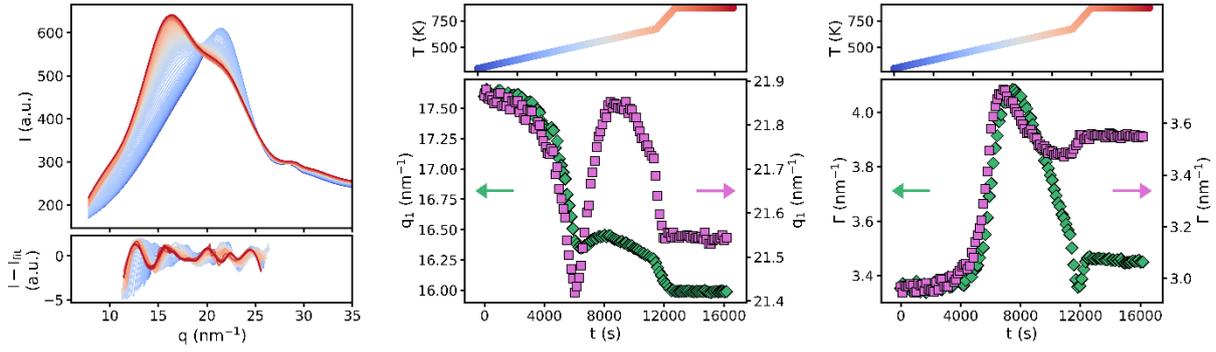

*Figure 4 – WAXS scattering curves for the NS3 glass (left), with the residuals from the modelling of the top 50% intensity of each curve by two gaussian functions. Corresponding evolution of the position (middle) and width (right) of each gaussian function. The color coding of the scattering curves and residuals corresponds to that of the temperature evolution.*

It is possible to model the top 50% of the FSDP in the NS3 glass throughout the full relaxation process with two Gaussian functions, as visible in the figure 4. The position (of the maximum) $q_1$ and width $\Gamma$ of each peak component are displayed in the figure 4. Both subpeaks shifts non-monotonously to lower $q$ values, which indicates a larger spacing in the electronic density correlations at this length scale, typically associated with a decrease of the macroscopic density [62,63], expected in the relaxation of densified glasses. At similar elapsed times of 8000-9000s, both sub peaks positions go through a local maximum in this overall decrease. This maximum is more marked in the high-$q$ (Na-rich glass like) component, reaching 1.5 nm$^{-1}$ against 0.02 nm$^{-1}$ in the low-$q$ component. This cannot be linked directly to a corresponding change macroscopic density change, as the connection between the FSDP and the density holds only *in the absence of major structural changes*, as shown in metallic systems with a similar thermo-mechanical history [64]. The same transient state is visible in the width $\Gamma$, which goes through a local maximum at the same elapsed time (temperature) of 8000s (565K), with a higher intensity in the low-q component (0.6 nm$^{-1}$) than in the high-q component (0.25 nm$^{-1}$). The same width goes through an a second local extremum at 12000s. This local extremum is reflected inversely in the peak intensity (figure S8), suggesting the possibility that the width and intensity parameters compensate each other in the fitting process, yielding an unphysical extremum.

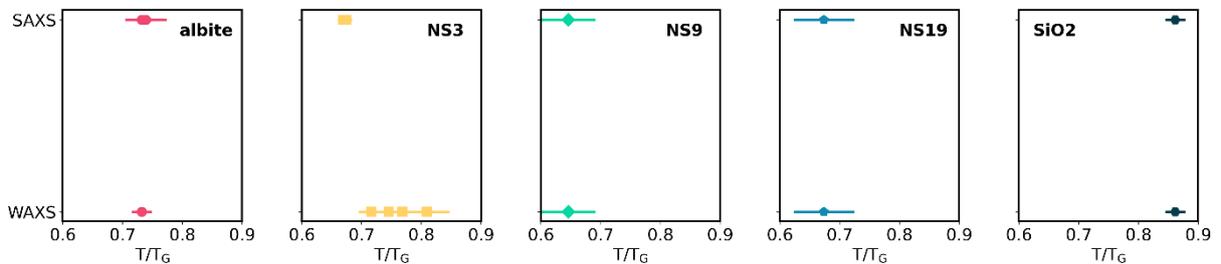

*Figure 5 - Temperature of the transient states from SAXS (Porod exponent and localized intensity) and WAXS (half width, and position for the NS3 glass only), scaled with respect to the glass transition temperature $T_G$. $T_G$ values for the binary glasses, albite and silica were taken from references [65], [66] and [67] respectively.*

The thermal stability of the HDA state is defined from the energy barrier separating it from the LDA state. This barrier is visible by all the transient feature identified in this study, so the height of the energy barrier can be estimated qualitatively from the temperature of this transient state with respect to the glass transition temperature at atmospheric pressure $T_G$, as shown in figure 5. The comparison between the different samples indicates that the stability of the HDA state is larger in fully polymerized glasses than in slightly depolymerized glasses, with the highest $T/T_G$ value in silica against NS19 and NS9. The intermediate value in albite suggest that in addition to depolymerization, the content of network modifiers also lowers the stability of the HDA state. Finally, nearly all samples show simultaneous transient states in the length scales covered by SAXS and WAXS. The NS3 sample differs, where both SAXS transient signatures (Porod exponent and localized intensity rise) appear at lower temperatures compared to the WAXS transient features, with no consistent order between the silica-like and high Na content components of the FSDP. The reason for this discrepancy between the different structural orders is unclear, but the higher $T/T_G$ values associated with the WAXS signatures suggests that the HDA state could regain stability for high Na content.

## 4 – Discussion

The evolution upon annealing of the structure of silica, binary $SiO_2$-$Na_2O$ and albite glasses previously densified from high temperature compression shows transient features in the length scales associated with small- and wide-angle scattering regimes. Qualitatively, groups of samples emerge: $SiO_2$, NS19, NS9 and albite show similar evolution of their FSDP, as opposed to that of the NS3 glass. In the small-angle scattering, $SiO_2$, NS19, NS9 are comparable with localized intensity variations, while NS3 and albite both show a spike emerging from the low-$q$ (Porod) intensity decrease. The similarity of $SiO_2$, NS19 and NS9 glasses indicate that the amount of Na is not sufficient to drastically change the structure of the silica network, which is the case in the NS3 glass. This is in line with previous reports that indicate a threshold of 20% $Na_2O$ in binary silicates in the evolution of elastic constants, Si coordination number and glass transition temperature [39]. The proximity of albite to $SiO_2$ at the length scale of the FSDP is also expected from the proximity between the melt structures at the level of the Si/Al-$O_{3/4}$ tetrahedra [68].

In this regard, the similarity of albite and NS3 in the nanometre scale is not straightforward, but our results indicate a strong resemblance between these samples.

The transient features in the SAXS intensity takes the form of a localized intensity maximum or a change in the slope of the Porod region. The variation of the Porod exponent indicates a transient change in the interface between Na-rich and Si-rich regions inside the glass structure. The direction of that spike toward lower values shows that this interface becomes more gradual with a higher degree of self-similarity [69]. In addition to the transient change in the Porod exponent, a local intensity maximum that forms of a single peak is visible for all glasses. In the Ornstein-Zernicke theory, a single peak in the scattering regime is consistent with an arrangement of hard spheres, whose shape is ideally Lorentzian [51,52]. The deviation from the Lorentzian shape likely comes from the unsharp boundaries of the density fluctuations, which leads to an additional $q$-dependence [70]. Thus, the observation of a single scattering peak indicates the apparition of fluctuations in the electronic density with a quasi-periodicity over a coherence length. This coherence length is determined by the width of the peak, up to 10 nm for silica.

Although there is a general consensus that the FSDP is arising from structural features of the MRO, and might even be representative of this MRO, no definitive interpretation of the actual density correlations giving rise to this scattering emerged, with multiple origins reported including tetrahedral ordering [71], chemical ordering of interstitial voids around cation-centred clusters [72,73] or remnants of crystal structures [74]. Because the most prominent structural feature of the MRO in amorphous silica corresponds to the rings formed by the $SiO_4$ tetrahedra (through their somital connections), the quasi-periodic density correlations corresponding to the FSDP is usually interpreted as an arrangement of such rings [75,76]. The different components within the FSDP even led to the suggestion of a quantitative connection suggested between the peak shape and the population of 4- 5- and 6- membered rings [77,78]. A similar interpretation of the density correlations related to voids around the network modifiers was suggested to account for the second peak in alkali-borates [58], providing a potential structural interpretation of the low-q peak visible in the FSDP of the NS3 glass. Recently, persistent homology emerged as a tool to probe directly the topological features (not necessarily supported by covalent bonds) of the MRO in oxide glasses [79,80]. Its application at atmospheric and high pressures showed that the statistics of ring sizes is not sufficient to capture the MRO and structural mechanisms behind compression and permanent densification, as the ring shape (compaction) matters at least equally [23,50,81,82], and direct observations of the density fluctuations behind the FSDP. Therefore, we consider here the FSDP as being representative of the MRO, but we limit ourselves to a qualitative interpretation. Even with this limited understanding of the underlying structural changes, the complex modifications of the FSDP through the different compositions during the temperature-assisted relaxations helps to understand the mechanism behind the potential polyamorphic transition. In particular, its evolution is not consistent with a simple density change or a variation of the coherence length of a structure that would remain qualitatively constant, as these would result in homogenous shifts and width variation respectively. The transient change observed at the level of the FSDP indicates specific structural change associated with a unique topology at the level of the intermediate range order that do not belong in the initial and end state of the overall relaxation. The time scale of the relaxation is order of magnitude larger than that available in simulations such as molecular dynamics, but the characterization of the particular topology of the transient state and the mechanism at play could be resolved with a Reverse Monte Carlo optimization of a glass structure based on diffraction data taken over the full $q$-range [23].

The observation of a transitory state implies that the pre and post transition structures belong to glassy states well separated by an energy barrier, hence belonging to two basins of the potential energy landscape. As the glass structure is inherited from the supercooled liquid state, we expect this transition to be properly defined in the phase diagram of the liquid state, although deep in the supercooled liquid region. This is in line with the temperatures reported for the transient state in figure 5 and joins the typical temperature range of liquid-liquid transitions in ionic, molecular, covalent and metallic systems [5,8,12,13], with only few exceptions [83–85]. As such, it reproduces the phenomenology observed in supercooled and glassy water, where the liquid-liquid (/glass-glass) transition (LLT) is well established [4,86] and the crossing of the liquid-liquid coexistence line or the Widom line (its continuation beyond the liquid-liquid critical point) is associated with a State of Strongest Heterogeneity (SSH) typical of the fluctuations expected at a phase transition or close to a critical point [87,88]. In water, this SSH appears in the SAXS intensity as a transient maximum at scattering vectors just above the Porod scattering region [89,90], as the one witnessed in this study for silicate glasses. The picture of developing fluctuations as a marker of a transition rather than a structural feature of a single liquid/glassy state also results from the correlation length of the transient SAXS maximum with values up to 10 nm in the case of silica. This is significantly larger than the correlation length of ≈ 1 nm in the glass or liquid structures obtained from the vanishing fluctuations in the positional (pair distribution function) [91,92] and rotational (4 point correlation function) [93] order. The lower $q$-position of the transient state in NS3 and albite glasses indicates a larger distance separating the volume of similar electronic densities, which in turn suggests a lower nucleation rate of the LDA state compared to the other samples. The nucleation rate of the LDA state seems correlated to the content of network modifiers, rather than to the polymerization degree of the glass structure.

One can only notice that the transition observed here in binary and ternary silicate glasses is not abrupt but smooth. There are two possible (and non-excluding) explanations for this, either the studied range of temperature (at 1 atm) is beyond the critical point where the transition is not well-defined anymore but appears as rise in fluctuations in the vicinity of the critical point, or it's a kinetic effect due the arrested nature of the glassy state. At the atomic length scale, a quantitative determination of the mechanism for the HDA→LDA transition is hindered by the low level of details present in the FSDP, but the observation of a transient state indicates a complex pathway involving a third structural state defined by unique topological features.

## 5 – Conclusion

This work uses x-ray scattering at small and wide angles to provide a structural description at the nanometre length scale of the relaxation of densified glass from hot compression. It includes the effects of network modifiers and polymerization by focusing on particular compositions in the ternary $SiO_2$-

Na$_2$O-Al$_2$O$_3$ system. As such, this study provides a novel view of the potential polyamorphism in amorphous silica, and an extension to silicate glasses. The analysis of a transient state in the SAXS regime reveals a water-like scenario where the fluctuations of the electronic density at large length scales (> 1 nm) correspond to the nucleation of a second state at a liquid-liquid transition or beyond but close to the associated critical point, all properly defined in the liquid state only. Therefore, our results suggest the presence of different liquid states and a potential critical point in the phase diagram of not only silica, but also in the 95SiO$_2$-5Na$_2$O, 90SiO$_2$-10Na$_2$O, 75SiO$_2$-30Na$_2$O and 75SiO$_2$-12.5Na$_2$O-12.5Al$_2$O$_3$ compositions. This indicates that the presence of polyamorphism is not linked on the polymerization degree in silicate glasses/liquids, even if the specific mechanisms at the atomic level do depend on the specific structure of the glass/liquid. The temperature scaling of the activation state in the HDA→LDA transition with respect to the glass transition temperature of each sample shows that the locus of the underlying LLT (or critical point) resides in the deeply undercooled regime, as for nearly all reports of LLTs in the literature.

## Acknowledgement

We gratefully acknowledge the ESRF (Grenoble, France) for providing access to beamtime, including experiments carried out on the BM26 beamline under the proposals MA3028 and MA3524. The authors wish to thank the whole team in the BM26 beamline in the ESRF, and especially Daniel Hermida Merino for the provision of in-house beamtime. High pressure experiments were performed at the PLECE-ILMTech platform.

# Supplementary Material for Multiscale Experimental Evidence of a Transitory State in the HDA-LDA Transition in Alumino-sodo-silicate Glasses

Additional information:

- Fig S1, S2, S3, S4: SAXS curves and intensity maps for pristine and densified NS19, NS9, NS3 and albite glasses.
- Fig S5: Characterization of the magnitude of the transient state in the SAXS intensity map of silica, NS19 and NS9 glasses.
- Fig S6: Comparison of temperature evolution of the SAXS intensity of the NS9 glass and Landau-Placzek ratio of a NS7.3 glass.
- Fig S7: Evolution of the FSDP maximum position of the silica, NS19, NS9 and albite glasses.
- Fig S8: Full modelling of the two components in the FSDP of the NS3 glass.

*Evolution of small angle scattering in densified and pristine glasses*

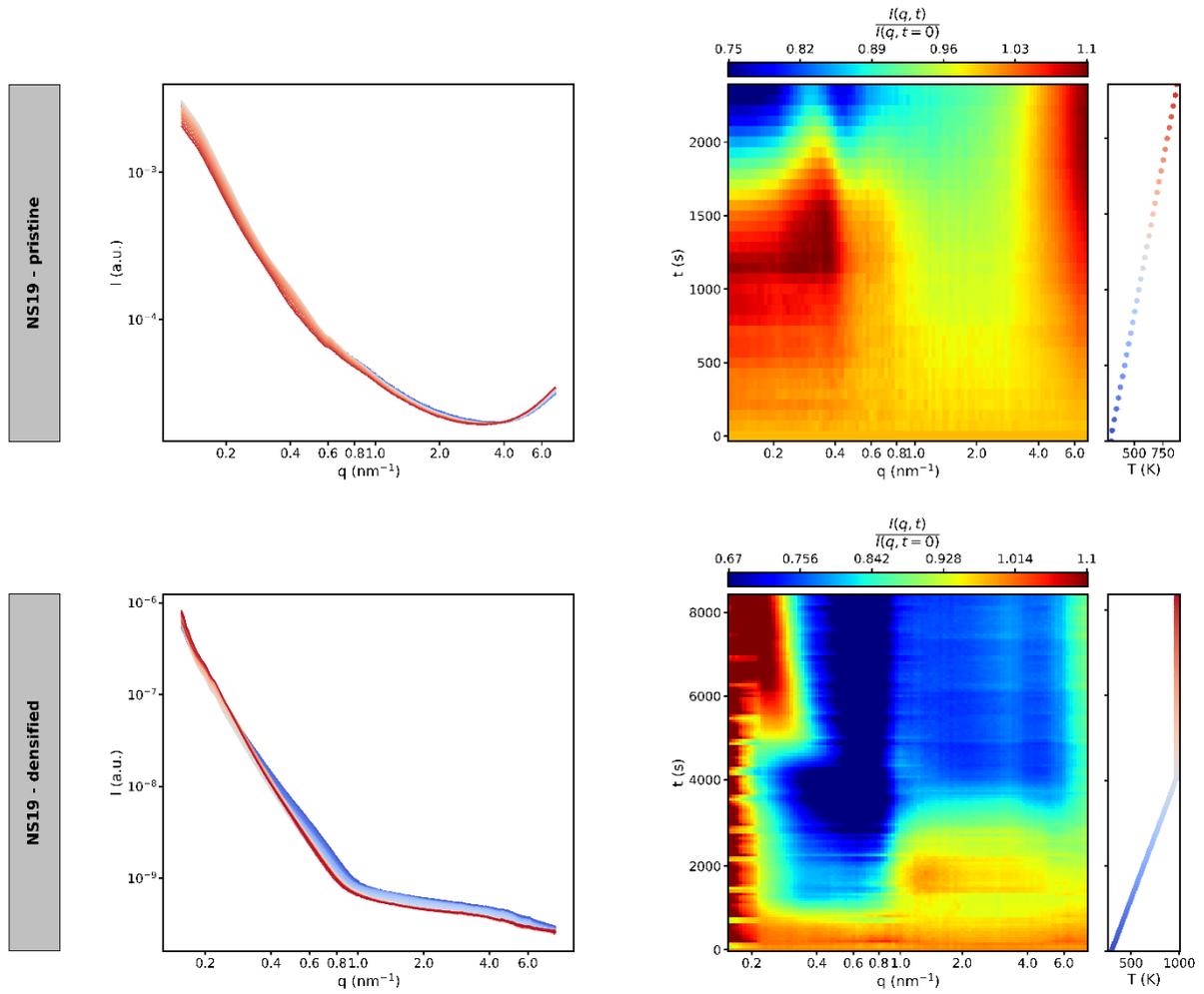

Figure S6 - X-ray small angle scattering curves and relative intensity map of NS19 pristine and densified glasses. The color coding of the scattering curve corresponds to that of the time-temperature evolution on the right side of the figure.

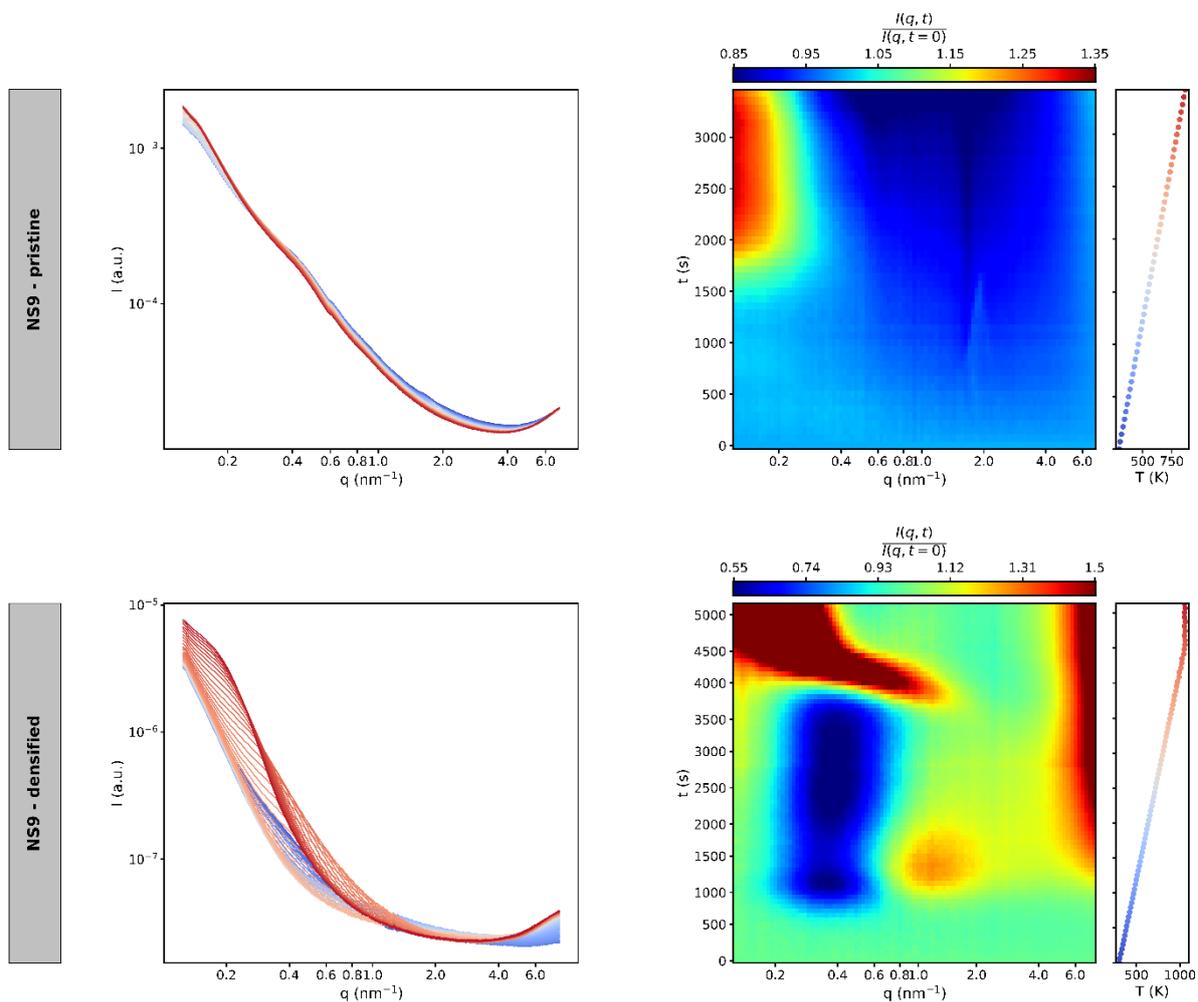

Figure S7 - X-ray small angle scattering curves and relative intensity map of NS19 pristine and densified glasses. The colour coding of the scattering curve corresponds to that of the time-temperature evolution on the right side of the figure.

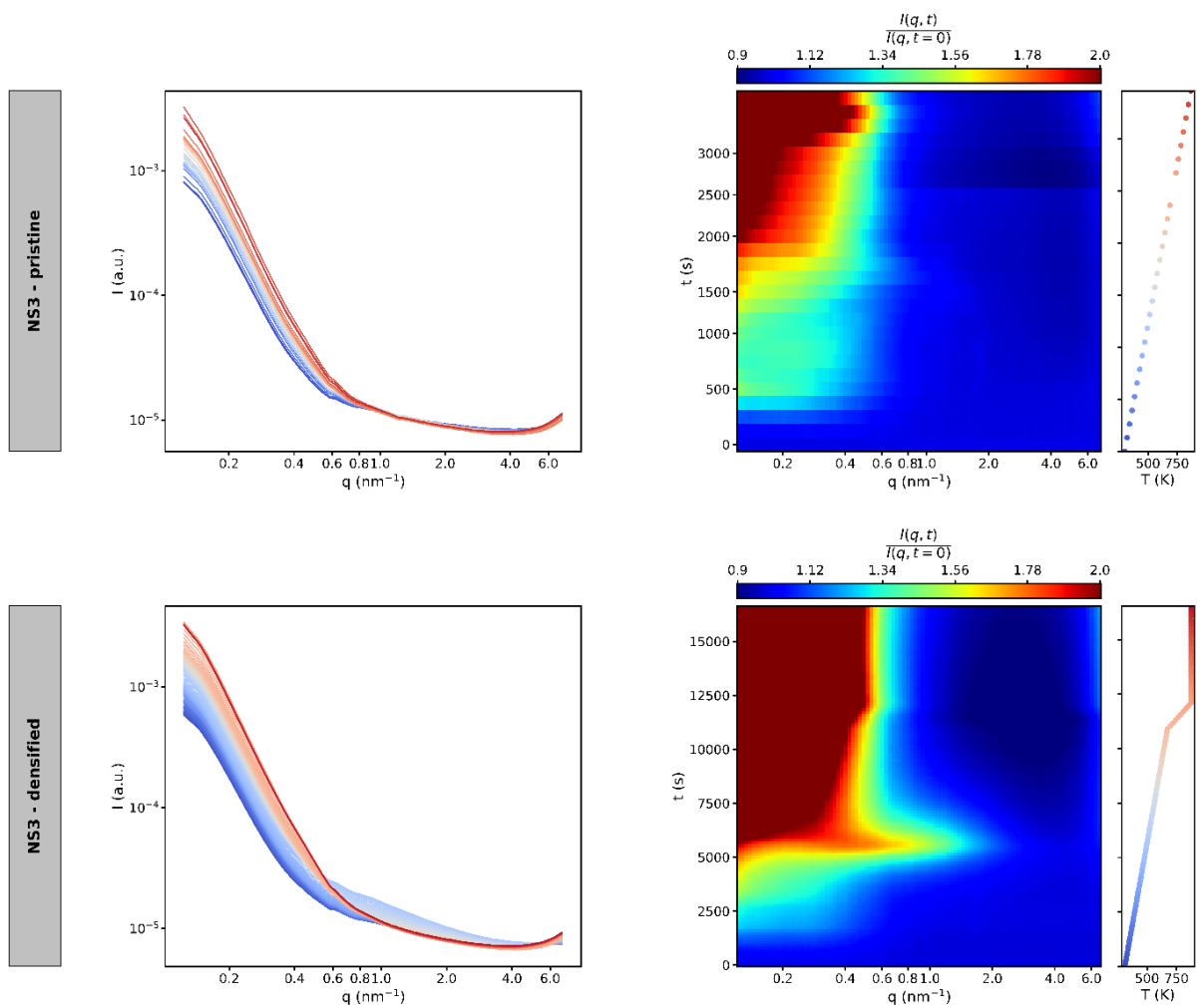

Figure S8 - X-ray small angle scattering curves and relative intensity map of NS19 pristine and densified glasses. The colour coding of the scattering curve corresponds to that of the time-temperature evolution on the right side of the figure.

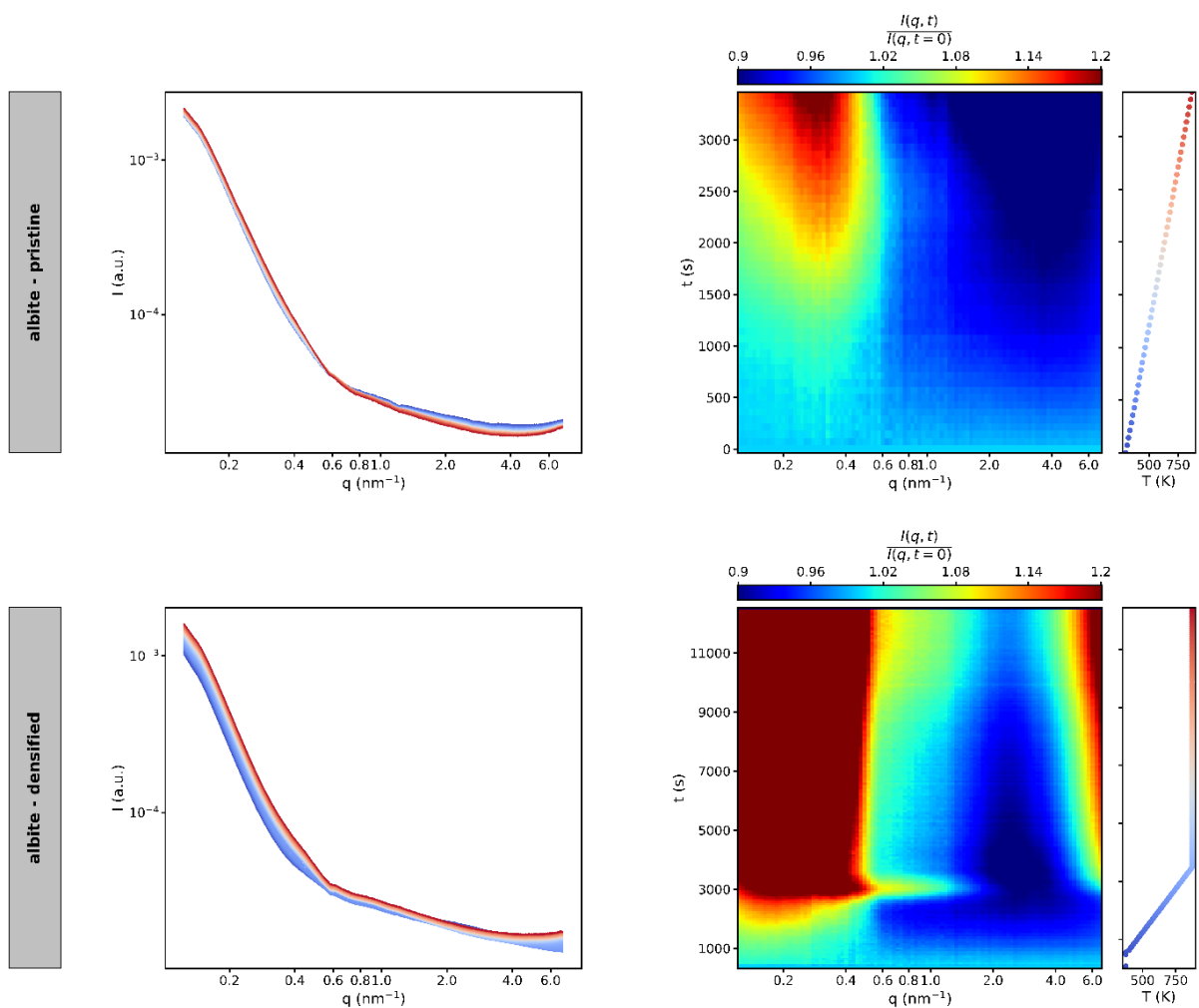

Figure S9 - X-ray small angle scattering curves and relative intensity map of NS19 pristine and densified glasses. The colour coding of the scattering curve corresponds to that of the time-temperature evolution on the right side of the figure.

*Quantification of the magnitude of the transient intensity maximum*

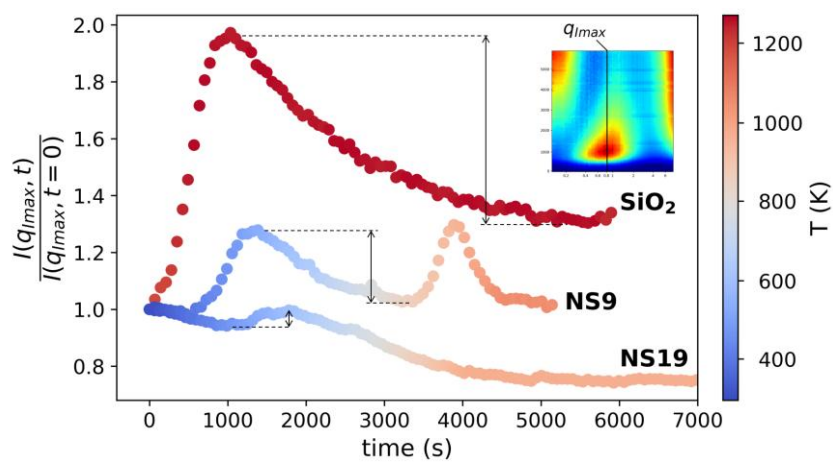

Figure S10 - Temporal evolution of the relative SAXS intensity at the scattering vector of the local intensity maximum, for the silica, NS19 and NS9 densified glasses.

The relative intensity evolution at the $q$-value of the transient state as a function of the elapsed time is represented in the figure S5 for the silica, NS19 and NS9 glasses. The corresponding temperature is colour coded. The choice of the $q$-value is illustrated in the inset for the silica glass. The magnitude of the transient state is characterized as the height of the local maximum relative to the highest minimum, as shown by the different two headed arrows and dashed lines. It yields values of 0.66±0.02, 0.06±0.02 and 0.25±0.02 in pure silica and in the NS19 and NS9 samples respectively.

*Selective identification of local intensity rise and chemical segregation: comparison of SAXS intensity and Landau-Placzek ratio in NS9 and NS7.3 glasses*

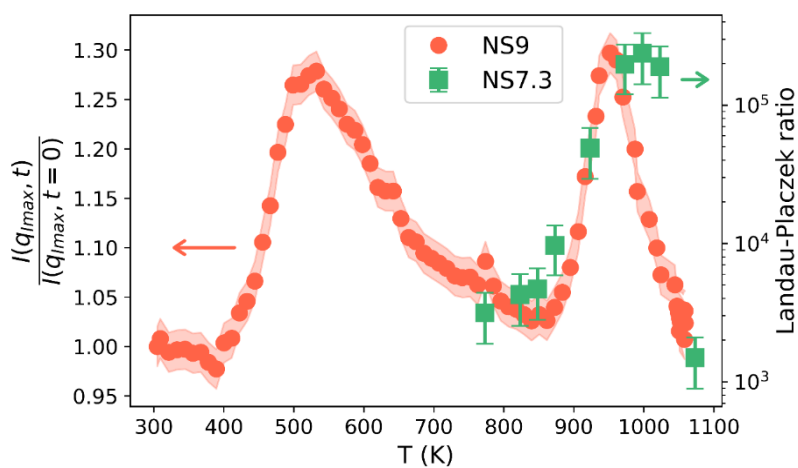

Figure S11 - Temperature evolution of the relative SAXS intensity in the NS9 glass, and temperature evolution of the Landu-Placzek ratio in a pristine NS7.3 glass.

## Width and position evolution of the FSDP in the silica, NS19, NS9 and albite glasses

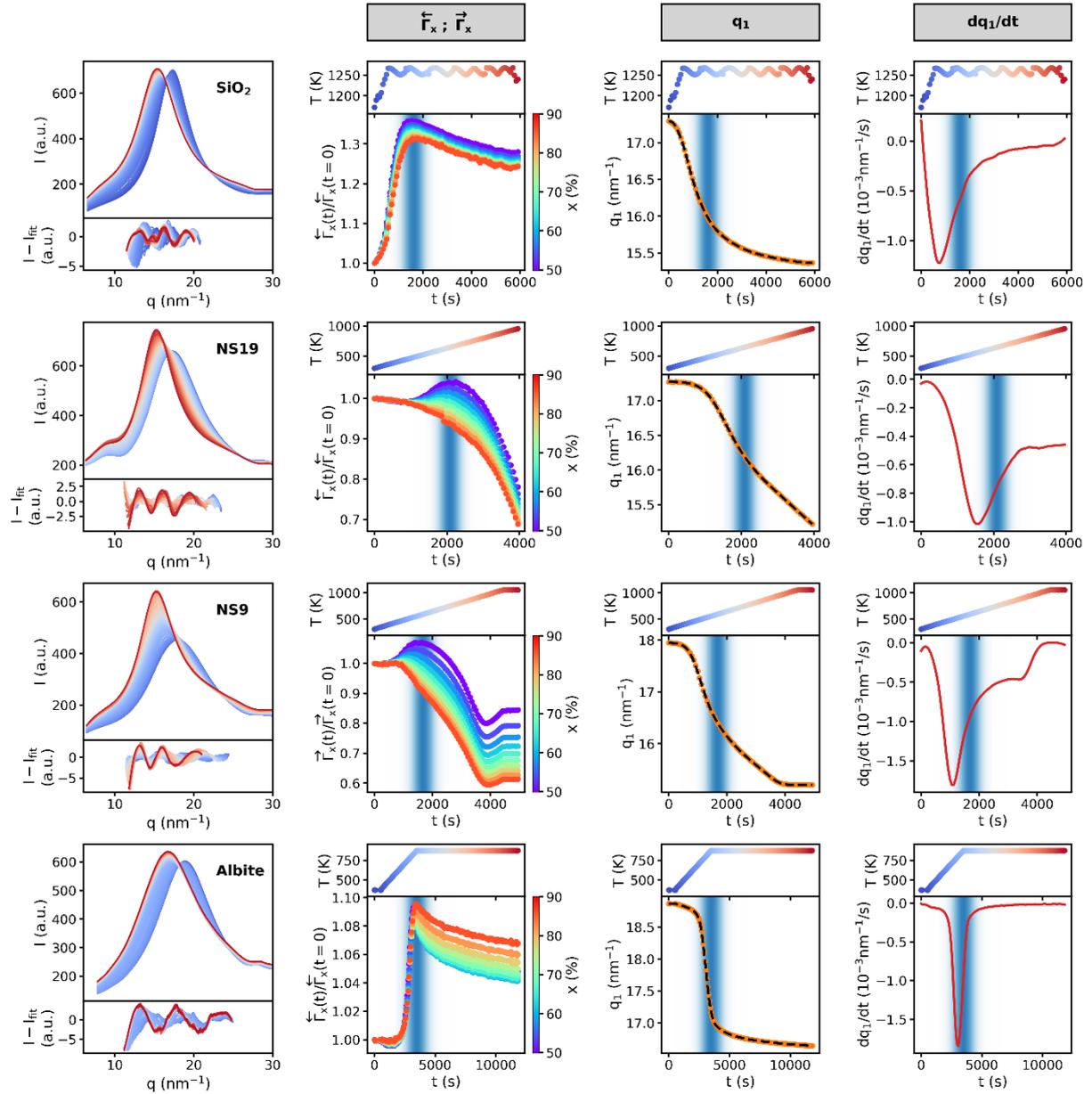

Figure S12 – FSDP, half width $\Gamma_x$, maximum position $q_1$ and first temporal derivative $dq_1/dt$ as a function of time. The secondary vertical panels in each row correspond to the fit residuals I-$I_{fit}$ and the temperature evolution. The colour coding of the scattering curves corresponds to that of the temperature evolution.

The figure S7 adds to the figure 3 of the main text, and represents the FSDP position, defined as the position of the maximum of the peak $q_1$, and its derivative as a function of time $dq_1/dt$. The half width showing the transient state is represented in the second panel of each row, and the time corresponding to the local maximum is represented by a blue shading, also plotted in the position and position derivative panels. The derivative is estimated from a Savitzky-Golay filter applied to the evolution of $q_1$, using a data window of 15 points. The value of the derivative in the first half window is less reliable, explaining the slight intensity maximum in the NS19 and NS9 glasses in the first 500s.

The position derivatives show a minimum for each glass, which indicate the maximum rate for the transition. This minimum appears before the time region of maximum width for all glass, although a near simultaneity is visible in the case of the albite glass. For all samples but silica, the derivative minimum takes place during the heating stage at constant rate, and the minimum in the temporal derivative translates directly to a minimum in the temperature derivative.

## *Modelling of the FSDP in the NS3 glass*

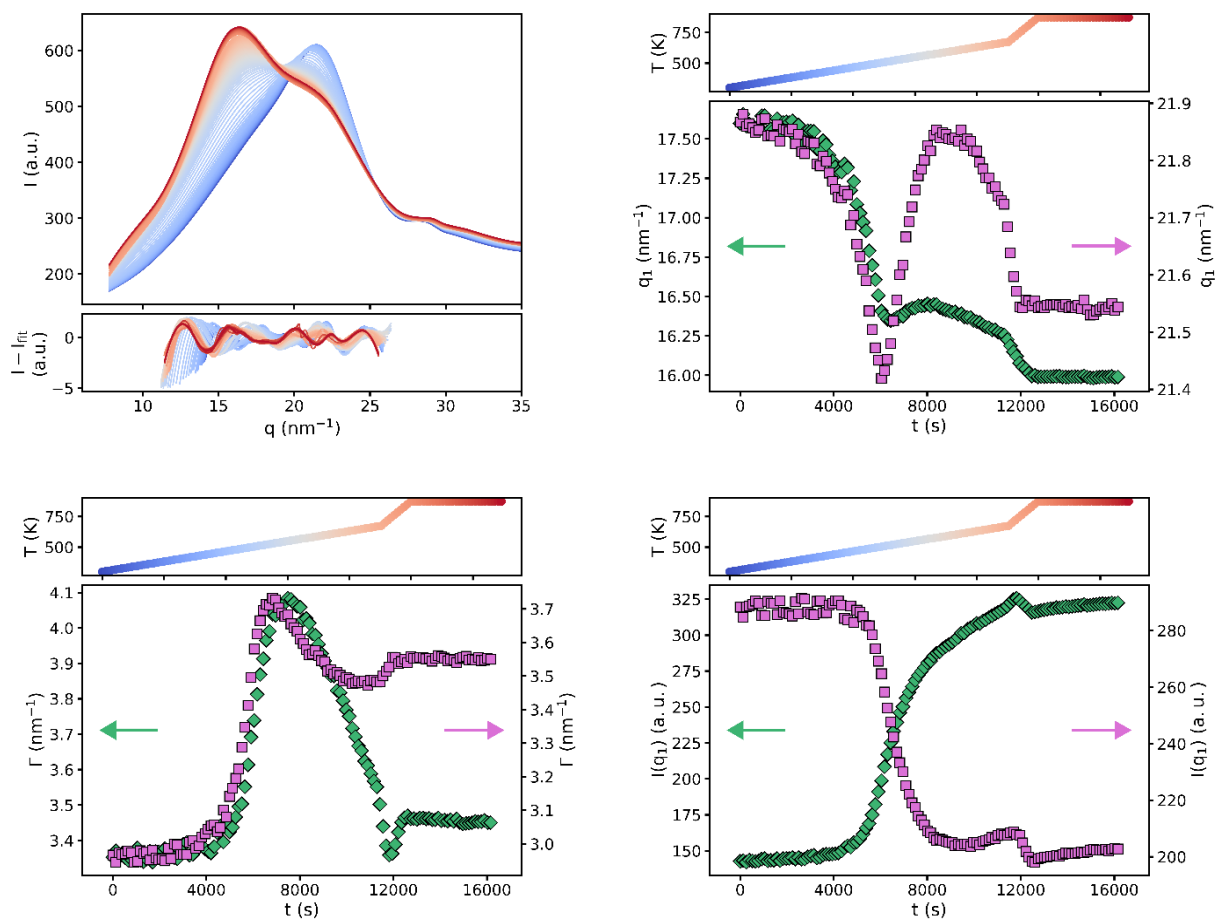

Figure S13 – FSDP, maximum position $q_1$, half width at half maximum $\Gamma$ and intensity maximum $I(q_1)$ of the two gaussian functions used to model the FSDP for the NS3 glass. The secondary vertical panels in each row correspond to the fit residuals I-I$_{fit}$ and the temperature evolution. The colour coding of the scattering curves corresponds to that of the temperature evolution.

The FSDP of the NS3 glass was modelled with two gaussian components. The residuals of this modelling of the upper half of the peak is shown just below the curve panel, and remains below 1% of the scattering intensity. The second local extremum visible on the width evolution at 12000s is reflected inversely on the peak heights, suggesting un unphysical compensation of the fit parameters.